\documentclass[journal]{IEEEtran}
\usepackage{graphicx}
\usepackage{subfigure}
\usepackage{subfloat}
\usepackage{booktabs}
\usepackage{picinpar}
\usepackage{amsfonts,amssymb}
\usepackage{mathrsfs}
\usepackage{amsmath}
\usepackage[numbers,sort&compress]{natbib}

\begin{document}
%
\title{\huge \bf Embedding Cryptographic Features in Compressive Sensing}

\author{Yushu Zhang,
        Kwok-Wo Wong,~\IEEEmembership{Senior Member,~IEEE},
        Di Xiao,~\IEEEmembership{Member,~IEEE},
        Leo Yu Zhang,
        and Ming Li
\thanks{Manuscript received Mar. 23, 2014. This work was supported by the Natural Science Foundation Project of CQ CSTC under Grant 2011jjjq40001.}
\thanks{Y. Zhang, D. Xiao and M. Li are with the College of Computer Science, Chongqing University, Chongqing, 400044 China (e-mail: yushuboshi@163.com; xiaodi\_cqu@hotmail.com; liming629@gmail.com) }
\thanks{K. Wong and L. Zhang are with the Department of Electronic Engineering, City University of Hong Kong, Kowloon, Hong Kong (e-mail: itkwwong@cityu.edu.hk; leocityu@gmail.com)}}


\vspace{-0.35in}

\markboth{}
{Shell \MakeLowercase{\textit{et al.}}: Bare Demo of IEEEtran.cls for Journals}
%

\maketitle

\begin{abstract}
Compressive sensing (CS) has been widely studied and applied in many fields. Recently, the way to perform secure compressive sensing (SCS) has become a topic of growing interest. The existing works on SCS usually take the sensing matrix as a key and the resultant security level is not evaluated in depth. They can only be considered as a preliminary exploration on SCS, but a concrete and operable encipher model is not given yet. In this paper, we are going to investigate SCS in a systematic way. The relationship between CS and symmetric-key cipher indicates some possible encryption models. To this end, we propose the two-level protection models (TLPM) for SCS which are developed from measurements taking and ``something else'', respectively. It is believed that these models will provide a new point of view and stimulate further research in both CS and cryptography. Specifically, an efficient and secure encryption scheme for parallel compressive sensing (PCS) is designed by embedding a two-layer protection in PCS using chaos. The first layer is undertaken by random permutation on a two-dimensional signal, which is proved to be an acceptable permutation with overwhelming probability. The other layer is to sample the permuted signal column by column with the same chaotic measurement matrix, which satisfies the restricted isometry property of PCS with overwhelming probability. Both the random permutation and the measurement matrix are constructed under the control of a chaotic system. Simulation results show that unlike the general joint compression and encryption schemes in which encryption always leads to the same or a lower compression ratio, the proposed approach of embedding encryption in PCS actually improves the compression performance. Besides, the proposed approach possesses high transmission robustness against additive Gaussian white noise and cropping attack.
\end{abstract}


\IEEEpeerreviewmaketitle

\vspace{0.2in}
\section{Introduction}

\subsection{Existing Work on Joint Compression and Encryption}

\IEEEPARstart{I}{}N digital signal processing, Shannon/Nyquist sampling theory is considered as the keystone in signal acquisition and reconstruction.  It governs the sampling process from the perspective of signal band limitedness. Traditionally, the sampled data are first compressed and then encrypted, if necessary. There is an increasing research interest on joint compression and encryption \cite{wu2005design, grangetto2006multimedia, wen2006binary, kim2007secure, bose2006novel,wong2010simultaneous,wong2008embedding,chen2011modified,yuen2012application,chen2012new} mostly by leveraging chaos theory and source coding schemes including arithmetic coding and Huffman coding. Wu and Kuo \cite{wu2005design} suggested two approaches for integrating encryption with multimedia compression systems, i.e., selective encryption and modified entropy coders with multiple statistical models. Both could be applied to the Huffman and the QM coders. Grangetto \emph{et al.} \cite{grangetto2006multimedia} introduced a randomized arithmetic coding paradigm which provides secrecy by introducing randomization in the coding process. Wen \emph{et al.} \cite{wen2006binary} described an interval splitting arithmetic coding in which a key known to both encoder and decoder is used to determine where the intervals are split prior to encoding each symbol. As a result, the traditional approach of using a single contiguous interval to encode a source symbol is relaxed. Kim \emph{et al.} \cite{kim2007secure} further analyzed the security of interval splitting arithmetic coding against chosen-plaintext attack and added a series of permutations at the input and output of the encoder. Furthermore, chaos has been employed in arithmetic coding for encryption purpose \cite{bose2006novel, wong2010simultaneous}. Bose and Pathak \cite{bose2006novel} presented a chaos-based adaptive arithmetic coding-encryption system. Unfortunately, this system has been found not decodable, since the current symbol to be encoded is swapped with a randomly selected symbol in the model prior to encoding, making it impossible for the decoder to mirror the encoder's operations \cite{zhou2008comments}. Observing that iterating a skew tent map reversely is equivalent to arithmetic coding, Wong \emph{et al.} \cite{wong2010simultaneous} proposed a simultaneous compression and encryption scheme in which the chaotic map for arithmetic coding is controlled by a secret key and keeps changing.

The aforementioned schemes are basically compression-orientated by embedding encryption in a compression algorithm. There are also a few reports like \cite{wong2008embedding, chen2011modified} on encryption-orientated schemes. Wong \emph{et al.} \cite{wong2008embedding} designed an algorithm for embedding compression in the Baptista-type chaotic cryptosystem and Chen \emph{et al.} \cite{chen2011modified} further improved its compression performance. Besides, there exist some chaos-based compression and encryption schemes aiming at multimedia data such as image \cite{yuen2012application} or video \cite{chen2012new}. These works are based on the traditional Shannon/Nyquist sampling theory. However, for a new sampling theory referred to as compressed sensing or compressive sampling (CS), it is worth to investigate how joint compression and encryption can be achieved. In other words, the topic to be investigated in this paper is how secrecy can be incorporated in CS theory.

Making use of the sparseness of natural signals, CS \cite{candes2006robust, donoho2006compressed, candes2008introduction, candes2006near} unifies sampling and compression to reduce the data acquisition and computational load at the encoder, at the cost of a higher computational complexity at the decoder. If the CS framework can integrate with certain cryptographic features for simultaneous sampling, compression and encryption, its application areas can be extended to, for example, limited-resource sensor and video surveillance. It has been suggested in \cite{candes2006near} that CS framework leads to an encryption scheme, where the sensing matrix can be considered as an encryption key. In recent years, there exist some pioneer works on secure compressive sensing (SCS) \cite{rachlin2008secrecy, orsdemir2008security, hossein2012security, agrawal2011secrecy, dautov2013establishing}. Rachlin and Baron \cite{rachlin2008secrecy} found that CS cannot achieve perfect secrecy but can guarantee computational secrecy. The definition of perfect secrecy \cite{shannon1949communication} requires that the occurrence probability of a message conditioned on the cryptogram is equal to the \emph{a priori} probability of the message, $P(X = x|Y = y) = P(X = x)$. Alternatively, the mutual information satisfies $I(X;Y) = 0$. In contrast to perfect secrecy, computational secrecy relies on the difficulty in solving a hard computational problem (e.g. NP-hard) at the computation resources available to the adversary. Orsdemir \emph{et al.} \cite{orsdemir2008security} investigated the security and robustness of employing a secret sensing matrix. They evaluated the security against brute force and structured attacks. The analyses indicate that the computational complexity of these attacks renders them infeasible in practice. In addition, this SCS method was found to have fair robustness against additive noise, making it a promising encryption technique for multimedia applications. Hossein \emph{et al.} \cite{hossein2012security} also addressed the perfect secrecy problem for the scenario that the measurement matrix as a key is known to both the sender and the receiver. Similar results have been obtained, as reported in \cite{rachlin2008secrecy}. It is shown that the Shannon perfect secrecy is, in general, not achievable by such a SCS method while a weaker sense of perfect secrecy may be achieved under certain conditions. Agrawal and Vishwanath \cite{agrawal2011secrecy} employed the CS framework to establish secure physical layer communication over a Wyner wiretap channel. They showed that CS can exploit channel asymmetry so that a message that is encoded as a sparse vector is decodable with high probability at the receiver while it is impossible to decode it with high probability by the eavesdropper. Dautov and Tsouri \cite{dautov2013establishing} proposed an encryption scheme where the sensing matrix is established using wireless physical layer security and linear feedback shift register with the corresponding $m$-sequences. It is shown that by using a Rician fading channel, the proposed scheme generates valid CS matrices while preventing access from an eavesdropper in close proximity to one of the legitimate participants.

\subsection{Our Contributions}
We will present two main contributions in this paper. On one hand, the CS-based encryption works \cite{rachlin2008secrecy, orsdemir2008security, hossein2012security, agrawal2011secrecy, dautov2013establishing}, only investigated SCS for the case that the measurement or sensing matrix serves as a key and the corresponding security is briefly evaluated. They have done a preliminary exploration on SCS but did not provide concrete and operable encipher models. Along this direction, we are going to investigate SCS in a systematic manner. The relationship between CS and symmetric cryptographic schemes indicates some possible encryption approaches. To this end, we propose the two-level protection models (TLPM) for SCS from two aspects of CS which are developed on measurements taking and ``something else'', respectively, as summarized in Table \ref{tab1}.
\begin{table*}[tb]
\caption{ Encipher models of SCS.}
\centering
\begin{tabular}{l c c}
\hline
Two aspects of CS & Encipher models and relevant references\\
\hline
1st-level: Measurements taking & Random projection \cite{yu2010compressive}  \\
&  Deterministic construction \cite{li2012deterministic} \\
& Structurally random matrices \cite{do2012fast}\\
& Multiclass encryption \cite{cambareri2013low}\\

2nd-level: ``Something else'' & Directional DCT during CS-based image coding \cite{zeng2008directional} \\
&  Side information in distributed CS \cite{deng2012robust}  \\
&  Permutation applied to parallel CS \cite{fang2013permutation}  \\
&  ... ... \\
\hline
\end{tabular}
\label{tab1}
\end{table*}

On the other hand, as discussed previously, the schemes embedding encryption in compression algorithms based on entropy coding or chaos theory suffer from the following drawbacks:

a. Low compression performance. Entropy coding based encryption always sacrifices the compression performance of entropy coding \cite{wen2006binary, kim2007secure, bose2006novel, wong2010simultaneous, yuen2012application, chen2012new} or at most maintains the same level of compression performance \cite{wu2005design, grangetto2006multimedia}.

b. Low robustness. The schemes \cite{wu2005design, grangetto2006multimedia, wen2006binary, kim2007secure, bose2006novel, wong2010simultaneous, yuen2012application, chen2012new} are too fragile to be applied in a noisy channel.

Specifically, we propose a SCS scheme embedding encryption in parallel compressed sensing (PCS) by chaos to overcome the above drawbacks. Our idea is inspired by two techniques \cite{fang2013permutation, frunzete2011compressive}: embed permutation in PCS and generate the measurement matrix by a tent map, respectively. The proposed scheme incorporates a two-layer protection into PCS under the control of chaos. The first layer is due to the random permutation on a two-dimensional (2D) sparse signal while the other samples the permuted signal column by column by using the same measurement matrix. Both the random permutation order and the measurement matrix are generated by a skew tent map. Our approach has the following superiorities:

a. Embedding random permutation based encryption enhances the compression performance.

b. The proposed approach possesses high transmission robustness against noise.

The rest of this paper is organized as follows. The next section introduces TLPM. In Section III, we review two recent techniques including embedding permutation in PCS and designing the measurement matrix by a tent map. By making use of these two techniques, a SCS scheme embedding a two-layer protection in PCS by chaos is proposed in Section IV, followed by simulation results in Section V and security analysis in Section VI. The last section concludes our work.

\vspace{-0.05in}
\section{Two-level Protection Models}

Suppose an $M$-dimensional signal ${\bf{f}} \in {\mathbb{R}^M}$ is expressed as

\begin{equation}
{\bf{f}} = \sum\nolimits_{i = 1}^M {{x_i}{{\bf{\psi }}_{\bf{i}}} = {\bf{\Psi x}}},
\end{equation}
which means that $\bf{f}$ could be sparsely represented in a certain domain by the transform matrix ${\bf{\Psi }}: = {\left[ {{{\bf{\psi }}_1},{\kern 1pt} {\kern 1pt} {{\bf{\psi }}_2},{\kern 1pt}  \cdots ,{\kern 1pt} {{\bf{\psi }}_M}} \right]}$ with each column vector ${{\bf{\psi }}_i} \in {\mathbb{R}^M}$, $i = 1,2, \ldots ,M$. We can say that $\bf{x}$ is exactly $k$-sparse if there are at most $k$ non-zero coefficients in the ${\bf{\Psi }}$ domain. Instead of sampling $\bf{x}$ directly, we take a small number of CS measurements. Let ${\bf{\Phi }}: = [{{\bf{\varphi }}_1},{\kern 1pt} {{\bf{\varphi }}_2},{\kern 1pt}  \cdots ,{{\bf{\varphi }}_M}]$ denote a $K \times M$ matrix with $K \ll M$. Then $K$ non-adaptive linear samples $\bf{y}$ can be obtained by

\begin{equation}
{\bf{y}} = {\bf{\Phi f}}.
\end{equation}
The resultant CS measurements $\bf{y}$ are used for the recovery of the original signal by solving the following convex optimization problem

\begin{align}
&\min {\left\| {\bf{x}} \right\|_1}\; s.t. \;{\bf{y = \Phi \Psi x}}\;\\
\nonumber
&(or\; in\; noisy\; situation:\; {\left\| {{\bf{\Phi \Psi x - y}}} \right\|_2} \le \varepsilon
)
\end{align}
to obtain ${\bf{\tilde f = \Psi x}}$.

\begin{figure*}[th]\centering
 \includegraphics[width=14 cm]{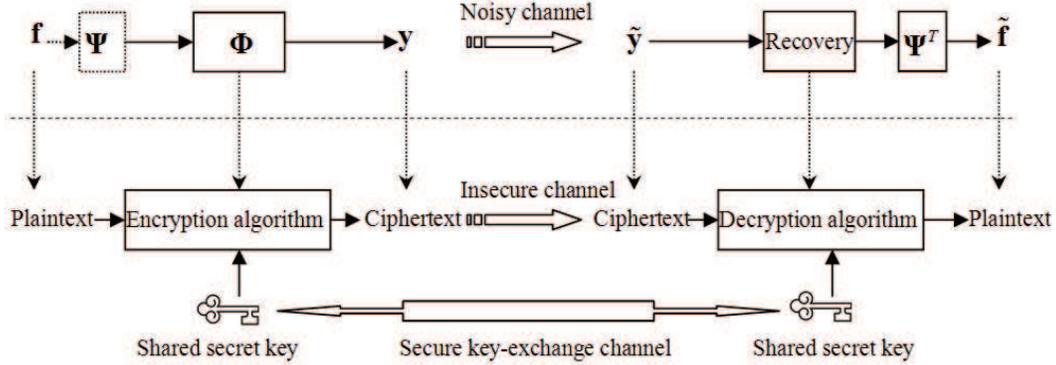}\\
  \caption{The relationship between CS and symmetric-key cipher.}
  \label{fig1}
\end{figure*}

The basic model for CS is shown in the upper half of Fig. \ref{fig1}, which includes two major aspects: measurements taking and signal recovery. From the perspective of symmetric-key cipher, measurements taking involves an encryption algorithm and signal recovery is associated with a decryption algorithm. The relationship between CS and symmetric cryptography indicates that some possible cryptographic features can be embedded in CS. To this end, we propose TLPM for SCS which are developed on measurements taking and ``something else'', respectively.

\subsection{1st-level Protection}
One of the central problems in CS framework is the selection of a proper measurement matrix ${\bf{\Phi }}$ satisfying the Restricted Isometry Property (RIP).

\noindent
\textbf{Definition 1:} \cite{candes2005decoding} Matrix ${\bf{\Phi }}$ satisfies the Restricted Isometry Property of order $s$ if there exists a constant ${\delta _s} \in \left[ {0,1} \right]$ such that

\begin{equation}
\left( {1 - {\delta _s}} \right)\left\| {\bf{x}} \right\|_2^2 \le \left\| {{\bf{\Phi x}}} \right\|_2^2 \le \left( {1 + {\delta _s}} \right)\left\| {\bf{x}} \right\|_2^2
\end{equation}
for all $s$-sparse signals ${\bf{x}}$ .

Cand\`{e}s and Tao \cite{candes2006near} proposed that a matrix following the Gaussian or Bernoulli distribution satisfies RIP with overwhelming probability at sparsity $s \le O\left( {{K \mathord{\left/ {\vphantom {K {\log M}}} \right. \kern-\nulldelimiterspace} {\log M}}} \right)$. The randomly selected Fourier basis also retains RIP with overwhelming probability, provided that the sparsity $s \le O\left( {{K \mathord{\left/  {\vphantom {K {{{\left( {\log M} \right)}^6}}}} \right. \kern-\nulldelimiterspace} {{{\left( {\log M} \right)}^6}}}} \right)$. In \cite{candes2006near}, it has been suggested that the sensing matrix can be used as an encryption key such that the CS framework leads to an encryption scheme.

Do \emph{et al.} \cite{do2012fast} proposed a framework called structurally random matrix (SRM), defined as

\begin{equation}
{\bf{\Phi }} = \sqrt {{K \over M}} {\bf{DFR}},
\end{equation}
where ${\bf{R}}\in \mathbb{R}^{M \times M}$ is either a uniform random permutation matrix or a diagonal random matrix whose diagonal entries ${R_{ii}}$ are i.i.d. Bernoulli random variables with identical distribution $P\left( {{R_{ii}} =  \pm 1} \right) = {1 \over 2}$. ${\bf{F}}\in \mathbb{R}^{M \times M}$ is an orthonormal matrix which is selected among popular fast computable transforms like DCT and ${\bf{D}}\in \mathbb{R}^{K \times M}$ represents a subsampling operator which selects a random subset of rows in the matrix ${\bf{FR}}$. The scalar coefficient $\sqrt {{K \over M}} $ is chosen to normalize the transform so as to ensure that the energy of the measurement vector is almost close to that of the input signal. This SRM can serve as a secret due to the fact that the random permutation matrix ${\bf{R}}$ is a common technique in classic encryption schemes.

Recently, Cambareri \emph{et al.} \cite{cambareri2013low} designed a novel multiclass encryption scheme based on perturbing the measurement matrix. The transmitter distributes the same encoded measurements to receivers with different privileges so that they are able to resconstruct the signal at various quality levels. Take the two-class situation as an example, the relationship between the two measurement matrices is formulated as

\begin{equation}
{{\bf{\Phi }}^{(1)}} = {{\bf{\Phi }}^{(0)}} + {\bf{\Delta \Phi }},
\end{equation}
where ${\bf{\Delta \Phi }}$ is an $c$-sparse perturbation matrix of entries

\begin{equation}
\Delta {\Phi _{i,j}} = \left\{ \begin{array}{l}
 0,{\kern 1pt} {\kern 1pt} {\kern 1pt} {\kern 1pt} {\kern 1pt} {\kern 1pt} {\kern 1pt} {\kern 1pt} {\kern 1pt} {\kern 1pt} {\kern 1pt} {\kern 1pt} {\kern 1pt} {\kern 1pt} {\kern 1pt} {\kern 1pt} {\kern 1pt} {\kern 1pt} {\kern 1pt} {\kern 1pt} {\kern 1pt} {\kern 1pt} {\kern 1pt} {\kern 1pt} {\kern 1pt} {\kern 1pt} {\kern 1pt} {\kern 1pt} {\kern 1pt} {\kern 1pt} {\kern 1pt} {\kern 1pt} {\kern 1pt} {\kern 1pt} {\kern 1pt} {\kern 1pt} {\kern 1pt} {\kern 1pt} {\kern 1pt} \left( {i,j} \right) \notin {C^{(0)}} \\
  - 2\Delta {\Phi _{i,j}},{\kern 1pt} {\kern 1pt} {\kern 1pt} {\kern 1pt} \left( {i,j} \right) \in {C^{(0)}}{\kern 1pt}  \\
 \end{array} \right.
\end{equation}
where ${C^{(0)}}$ is a subset of $c < KM$ entries chosen at random for each ${{\bf{\Phi }}^{(0)}}$ with density ${c \mathord{\left/
{\vphantom {c {KM}}} \right. \kern-\nulldelimiterspace} {KM}}$. A first-class user knowing the complete sampling matrix ${{\bf{\Phi }}^{(1)}}$
is able to exactly recover while a second-class user only having the knowledge of ${{\bf{\Phi }}^{(0)}}$ is instead subject to an equivalent non-white noise term ${\bf{\varepsilon }} = {\bf{\Delta \Phi x}}$ because of the true sampling ${\bf{y}} = {{\bf{\Phi }}^{(1)}}{\bf{x}}$.

Li \emph{et al.} \cite{li2012deterministic} introduced a deterministic construction of sensing matrix via algebraic curves over finite fields, which is a natural generalization of DeVore's construction \cite{devore2007deterministic} using polynomials over finite fields. The diversity of algebraic curves provides numerous choices for the sensing matrices, i.e., more choices of key in the encryption scheme, which may be valuable for the potential use of the sensing matrix for cryptographic purpose. Besides, it has been investigated that chaotic sequences can be employed to construct the sensing matrix \cite{yu2010compressive, frunzete2011compressive}, which will be further discussed in Section III and used in the proposed encryption scheme in Section IV.

\subsection{2nd-level Protection}

Apart from the above encryption mode on measurement taking, in each practical application of CS, some other cryptographic measures can be incorporated into other aspects that we term ``something else'' according to the specific situation.

Take CS-based image coding \cite{deng2012robust} as an example. Deng \emph{et al.} \cite{deng2012robust} proposed an alternative image coding paradigm with a number of descriptions based upon CS for high packet loss transmission. After a 2D DWT is applied for sparse representation, DWT coefficients are re-sampled towards equal importance of information instead. At the decoder side, two different CS recovery algorithms are developed for the low-frequency and the high-frequency subbands, respectively, by fully exploiting the intra-scale and inter-scale correlation of multiscale DWT. Experimental results showed this CS-based image codec is much more robust in lossy channels. During the recovery of scaling coefficients, spatial Gabor filters with different frequencies and orientations \cite{zeng2008directional} are utilized to extract the dominant orientation of structures in each block. The Gabor filter kernels with eight directions: ${{j \cdot \pi } \mathord{\left/ {\vphantom {{j \cdot \pi } 8}} \right. \kern-\nulldelimiterspace} 8}$, $j = 0,1, \cdots ,7$,   are used to output a response with respective to the filter kernel for each scaling coefficient. The largest response value is selected as the representative orientation corresponding to one of the eight intra predictors, which is sent to the decoder as side information. Such side information can serve as a key. Side information is commonly used in distributed coding \cite{slepian1973noiseless, wyner1976rate, pradhan2003distributed} and its role as a key has been discussed in \cite{johnson2004compressing, schonberg2008toward, klinc2012compression}. This is an interesting research branch on compressing encrypted data, which may lead to a new research direction of compressing encrypted data in distributed CS \cite{do2009distributed}.

Fang \emph{et al.} \cite{fang2013permutation} relaxed the RIP for 2D sparse signals by permutation in PCS, in which the 2D signal is sampled column by column using the same sensing matrix. In particular, the zigzag-scan permutation as a so-called acceptable permutation is applied to the 2D signal, the corresponding sensing matrix has a smaller required order of RIP condition; therefore, storage and computation requirements are further reduced. This idea of permutation inspires us whether the zigzag-scan permutation extended to random permutation can provide cryptographic features in PCS. This topic will be discussed in the following two sections.

\vspace{-0.05in}
\section{Two Techniques}
The background of TLPM has been given in the previous section. In the following sections, an encryption scheme for PCS will be designed by applying two techniques, random permutation and chaotic measurement matrix, in the 2nd-level and 1st-level protection models, respectively.

\subsection{Embedding Random Permutation in Parallel Compressive Sensing}
Traditionally, a multidimensional signal needs to be reshaped into an 1D signal prior to sampling using CS. Nevertheless, such a transformation makes the required size of the sensing matrix dramatically large and increases the storage and computational complexity significantly. To solve this problem, Fang \emph{et al.} proposed a novel solution \cite{fang2013permutation}, referred to as parallel compressed sensing, which reshapes the multidimensional signal into a 2D signal and samples the latter column by column with the same sensing matrix. Moreover, a so-called acceptable permutation can effectively relax the RIP for PCS.

\noindent
\textbf{Definition 2:} \cite{fang2013permutation} For a 2D sparse signal ${\bf{X}}$ with sparsity vector ${\bf{s}} = \left[ {{s_1},{s_2}, \cdots ,{s_N}} \right]$ satisfying ${\left\| {\bf{s}} \right\|_1} = s$, where ${s_j}$ is the sparsity level of the $j$-th column of ${\bf{X}}$, a permutation ${\bf{P}(\bullet)}$ is called acceptable for ${\bf{X}}$  if  the Chebyshev norm of the sparsity vector of $\bf{P}\left( {\bf{X}} \right)$ is smaller than ${\left\| {\bf{s}} \right\|_\infty }$ of ${\bf{X}}$.

When a 2D ${\bf{s}}$-sparse signal is exactly reconstructed by using PCS, a sufficient condition is given by the following lemma.

\noindent
\textbf{Lemma 1:} \cite{fang2013permutation} Consider a 2D ${\bf{s}}$-sparse signal ${\bf{X}}$, if the RIP of order ${\left\| {\bf{s}} \right\|_\infty }$ holds for the sensing matrix ${\bf{\Phi }}$, i.e., ${\delta _{2{{\left\| {\bf{s}} \right\|}_\infty }}} < \sqrt 2  - 1$, then ${\bf{X}}$ can be exactly reconstructed using PCS scheme.

This lemma implies that with respect to PCS, the RIP requirement of the sensing matrix ${\bf{\Phi}}$ at a given reconstruction quality is related to ${\left\| {\bf{s}} \right\|_\infty }$. A zigzag-scan permutation is considered acceptable in relaxing the RIP condition before using the PCS \cite{fang2013permutation}, but it is tailored to the sparse signal following a layer model. We generalize the permutation for the 2D sparse signal whose distribution is unknown in advance. Assume that ${\bf{P}(\bullet)}$ is a random permutation operation, then ${{\bf{X}}^ * } = {\bf{P}}\left( {\bf{X}} \right)$, where ${{\bf{X}}^ * } \in \mathbb{R}^{M \times N}$ is a permuted 2D signal with sparsity vector ${{\bf{s}}^ * }$. Observing the relationship between random and acceptable permutations, we have the following theorem.

\noindent
\textbf{Theorem 1:} For a 2D sparse signal ${\bf{X}}$, if the distribution of the sparsity level in each column is not sufficiently uniform, then the random permutation ${\bf{P}(\bullet)}$ is an acceptable permutation with overwhelming probability

\begin{equation}
P\left\{ {{\bf{P}}\left( \bullet \right){\kern 1pt} {\kern 1pt} {\kern 1pt} {\kern 1pt} is{\kern 1pt} {\kern 1pt} acceptable} \right\} \buildrel\textstyle.\over= 1 - \frac{1}{{{N^{\left\lceil {{s \mathord{\left/
 {\vphantom {s N}} \right.
 \kern-\nulldelimiterspace} N}} \right\rceil }}}}.
\end{equation}

The proof of this theorem can be found in Appendix A.

\begin{figure*}[th]\centering
 \includegraphics[width=14 cm]{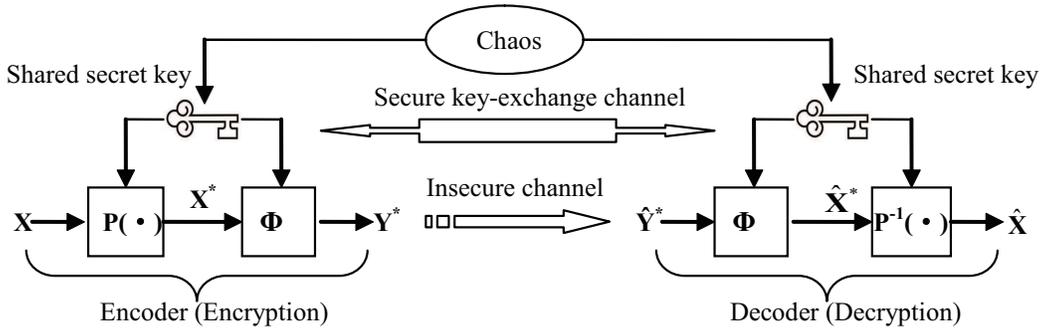}\\
  \caption{ A schematic diagram of the proposed approach.}
 \label{fig2}
\end{figure*}

\subsection{Chaotic Measurement Matrix}
For a random sensing matrix, its storage and transmission require a lot of space and energy. Thus, it is preferable to generate and handle the sensing matrix by one or more seed keys only. Yu \emph{et al.} \cite{yu2010compressive} proposed to construct the sensing matrix using chaotic sequence in a trivial manner and proved that the RIP of this kind of matrix is guaranteed with overwhelming probability, providing that the sparsity $s \le O\left( {{K \mathord{\left/ {\vphantom {K {\log \left( {{M \mathord{\left/ {\vphantom {M s}} \right. \kern-\nulldelimiterspace} s}} \right)}}} \right. \kern-\nulldelimiterspace} {\log \left( {{M \mathord{\left/ {\vphantom {M s}} \right. \kern-\nulldelimiterspace} s}} \right)}}} \right)$. They generated a sampled Logistic sequence $X\left( {d,l,{z_0}} \right)$, which has been regularized, with sampling distance $d$, length $l = K \times M$ and initial condition ${z_0}$. Then a matrix ${\bf{\Phi }}$ is created from this sequence column by column, denoted as

\begin{equation}
{\bf{\Phi }} = \sqrt {\frac{2}{K}} \left( {\begin{array}{*{20}{c}}
   {{z_0}} & {{z_K}} &  \cdots  & {{z_{K(M - 1)}}}  \\
   {{z_1}} & {{z_{K + 1}}} &  \cdots  & {{z_{K(M - 1) + 1}}}  \\
    \vdots  &  \vdots  &  \ddots  &  \vdots   \\
   {{z_{K - 1}}} & {{z_{2K - 1}}} &  \cdots  & {{z_{KM - 1}}}  \\
\end{array}} \right)
\end{equation}
where the scalar $\sqrt {{2 \mathord{\left/ {\vphantom {2 K}} \right. \kern-\nulldelimiterspace} K}}$ is for normalization purpose. One can take the initial condition ${z_0}$ as a key, since different sensing matrices are obtained from different initial conditions. The adoption of chaos can further enhance the security due to its pseudo-random behavior and high sensitivity to the initial condition.

Frunzete \emph{et al.} \cite{frunzete2011compressive} further constructed chaotic measurement matrix by introducing the one-dimensional skew tent map given by

\begin{equation}
z\left( {k + 1} \right) = T\left[ {z\left( k \right);\mu } \right] = \left\{ \begin{array}{l}
 \frac{{z\left( k \right)}}{\mu },{\kern 1pt} {\kern 1pt} {\kern 1pt} {\kern 1pt} {\kern 1pt} {\kern 1pt} {\kern 1pt} {\kern 1pt} {\kern 1pt} {\kern 1pt} {\kern 1pt} {\kern 1pt} {\kern 1pt} {\kern 1pt} {\kern 1pt} {\kern 1pt} {\kern 1pt} {\kern 1pt} {\kern 1pt} if{\kern 1pt} {\kern 1pt} {\kern 1pt} 0 < z\left( k \right) < \mu  \\
 \frac{{1 - z\left( k \right)}}{{1 - \mu }},{\kern 1pt} {\kern 1pt} {\kern 1pt} {\kern 1pt} if{\kern 1pt} {\kern 1pt} {\kern 1pt} \mu  \le z\left( k \right) < 1 \\
 \end{array} \right.
\end{equation}
where the control parameter $\mu  \in \left( {0,1} \right)$ and the initial state $z\left( 0 \right) \in \left( {0,1} \right)$. In terms of RIP, Frunzete \emph{et. al.} \cite{frunzete2011compressive} proved the following theorem.

\noindent
\textbf{Theorem 2:} \cite{frunzete2011compressive} A chaotic sequence ${\bf{{\bf{\Phi }}}} \in \mathbb{R}^{K \times M}$ constructed by the skew tent map satisfies RIP for constant ${\delta _s} > 0$ with overwhelming probability providing that $s \le O\left( {{K \mathord{\left/ {\vphantom {K {\log \left( {{M \mathord{\left/  {\vphantom {M s}} \right. \kern-\nulldelimiterspace} s}} \right)}}} \right. \kern-\nulldelimiterspace} {\log \left( {{M \mathord{\left/ {\vphantom {M s}} \right. \kern-\nulldelimiterspace} s}} \right)}}} \right)$.

For PCS, we can immediately infer the result described below.

\noindent
\textbf{Corollary 1:} If the sensing matrix ${\bf{\Phi}} \in \mathbb{R}^{K \times M}$ is constructed by a chaotic sequence with

\begin{equation}
K \ge C \cdot {\left\| {\bf{s}} \right\|_\infty }\log \left( {{M \mathord{\left/
 {\vphantom {M {{{\left\| {\bf{s}} \right\|}_\infty }}}} \right.
 \kern-\nulldelimiterspace} {{{\left\| {\bf{s}} \right\|}_\infty }}}} \right)
\end{equation}
for some constant $C$, then it will satisfy the RIP of order ${\left\| {\bf{s}} \right\|_\infty }$ with overwhelming probability.

Unlike the Logistic map, the probability density function of skew tent map follows the uniform distribution, which is more immune against statistical attacks in cryptographic applications. As a result, the skew tent map is chosen in the permutation operation and the construction of the measurement matrix.

\begin{figure*}[th]
\centering
\subfigure[]{
\begin{minipage}[t]{0.2\linewidth}
\includegraphics[width=\textwidth]{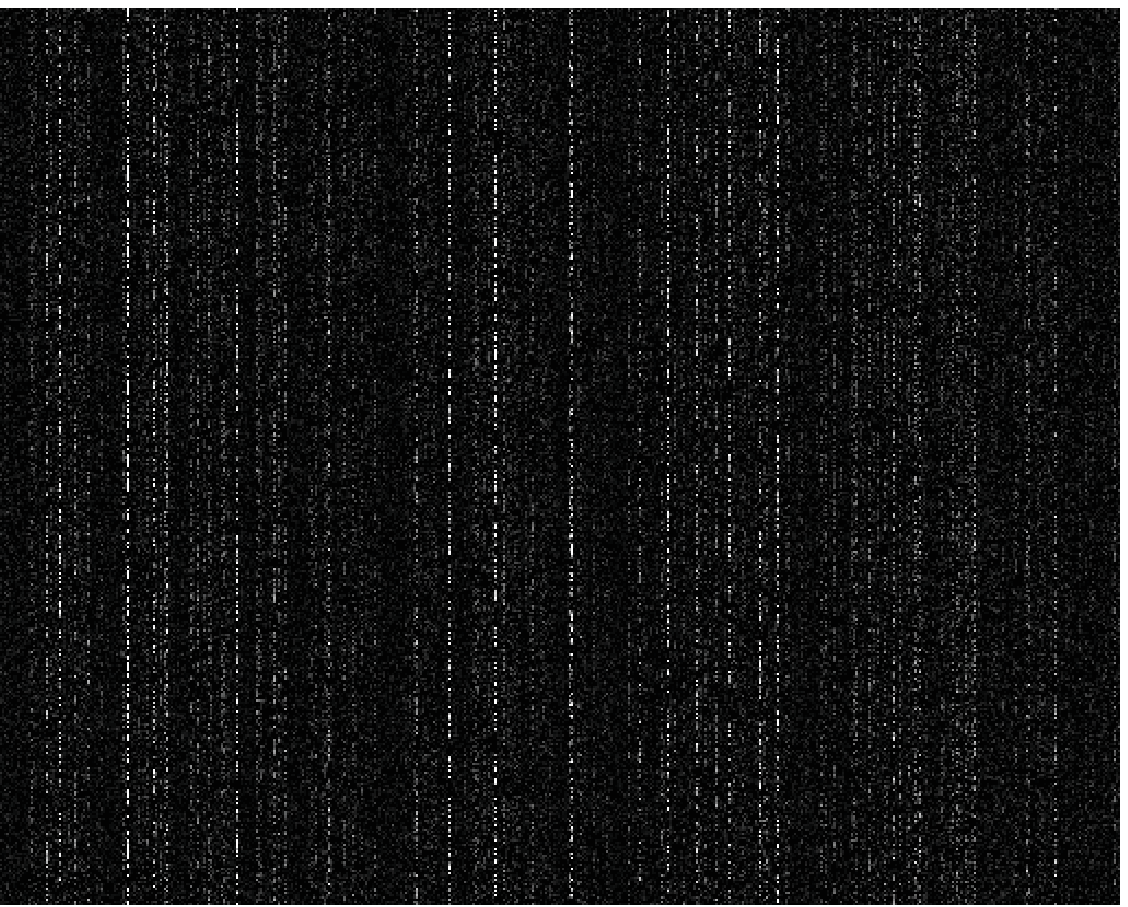}
\end{minipage}}%
\subfigure[]{
\begin{minipage}[t]{0.2\linewidth}
\includegraphics[width=\textwidth]{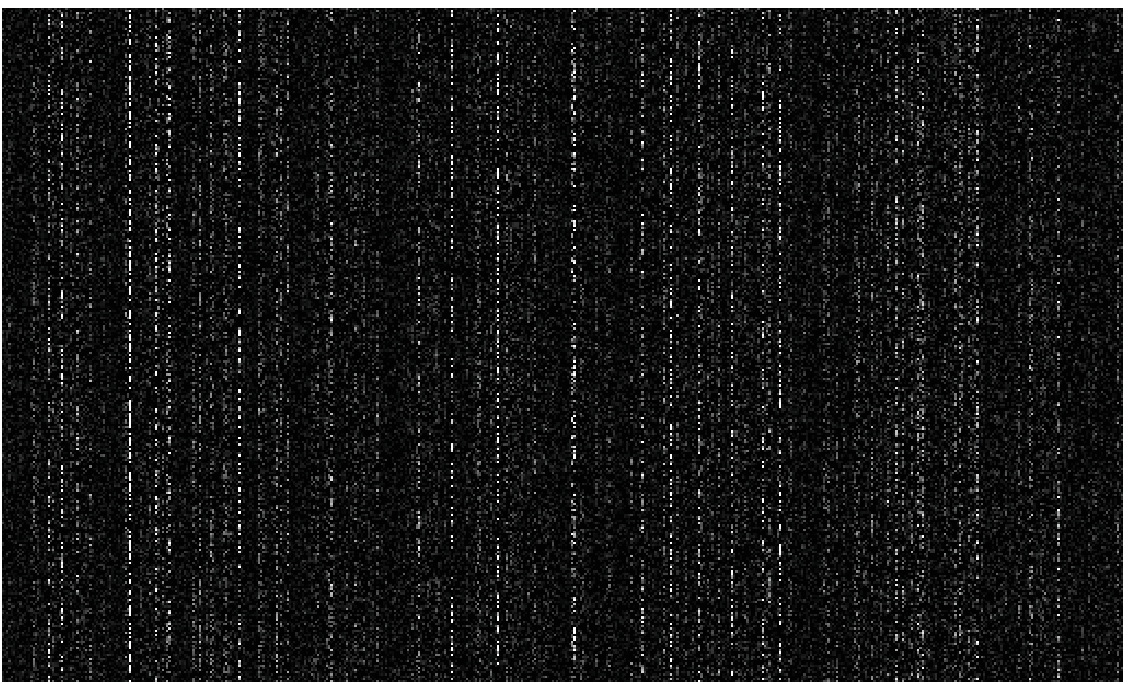}
\end{minipage}}
\subfigure[]{
\begin{minipage}[t]{0.2\linewidth}
\includegraphics[width=\textwidth]{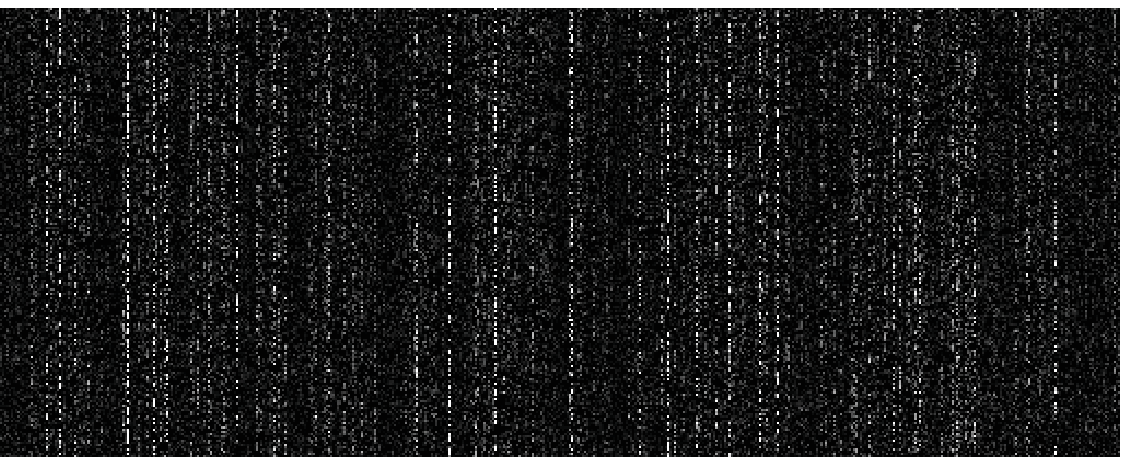}
\end{minipage}}%
\subfigure[]{
\begin{minipage}[t]{0.2\linewidth}
\includegraphics[width=\textwidth]{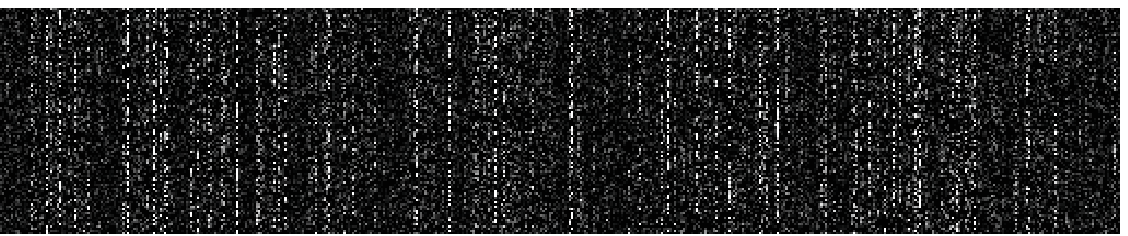}
\end{minipage}}
\caption{\small Four encoded images at different $CR$s. (a) $CR = 0.8$; (b) $CR = 0.6$; (c) $CR = 0.4$; (d) $CR = 0.2$.}
\label{fig3}
\end{figure*}

\begin{figure*}[th]
\centering
\subfigure[]{
\begin{minipage}[t]{0.2\linewidth}
\includegraphics[width=\textwidth]{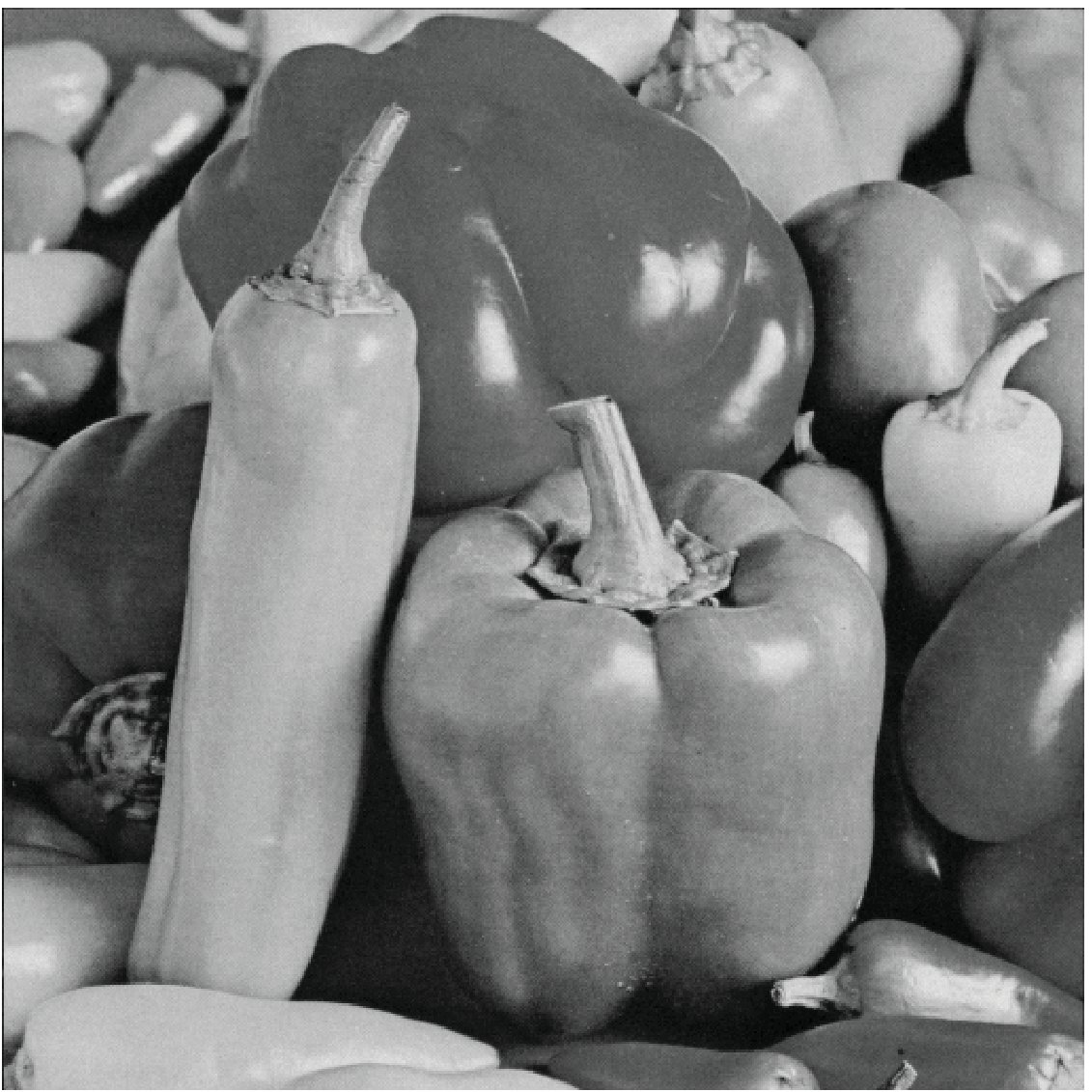}
\end{minipage}}%
\subfigure[]{
\begin{minipage}[t]{0.2\linewidth}
\includegraphics[width=\textwidth]{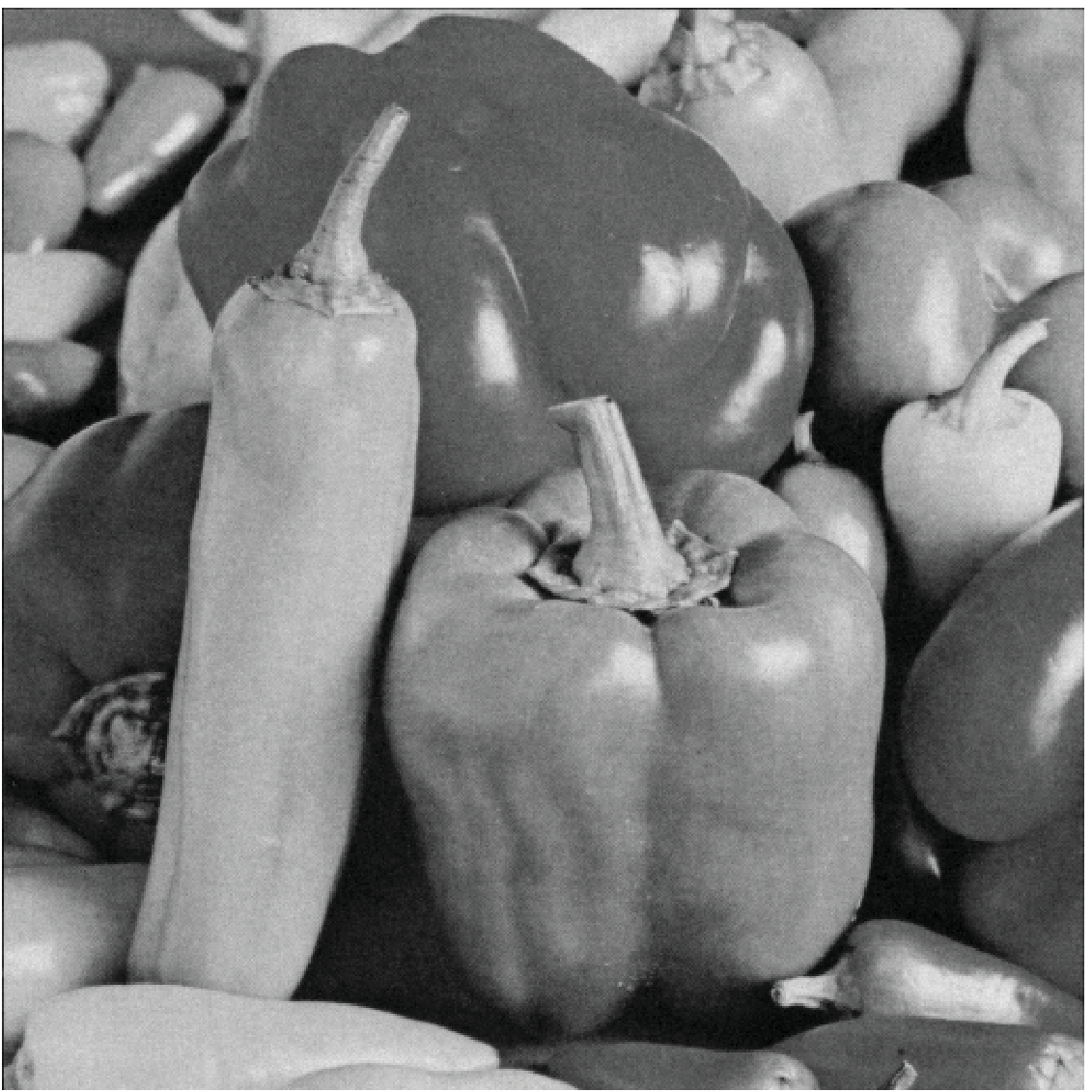}
\end{minipage}}
\subfigure[]{
\begin{minipage}[t]{0.2\linewidth}
\includegraphics[width=\textwidth]{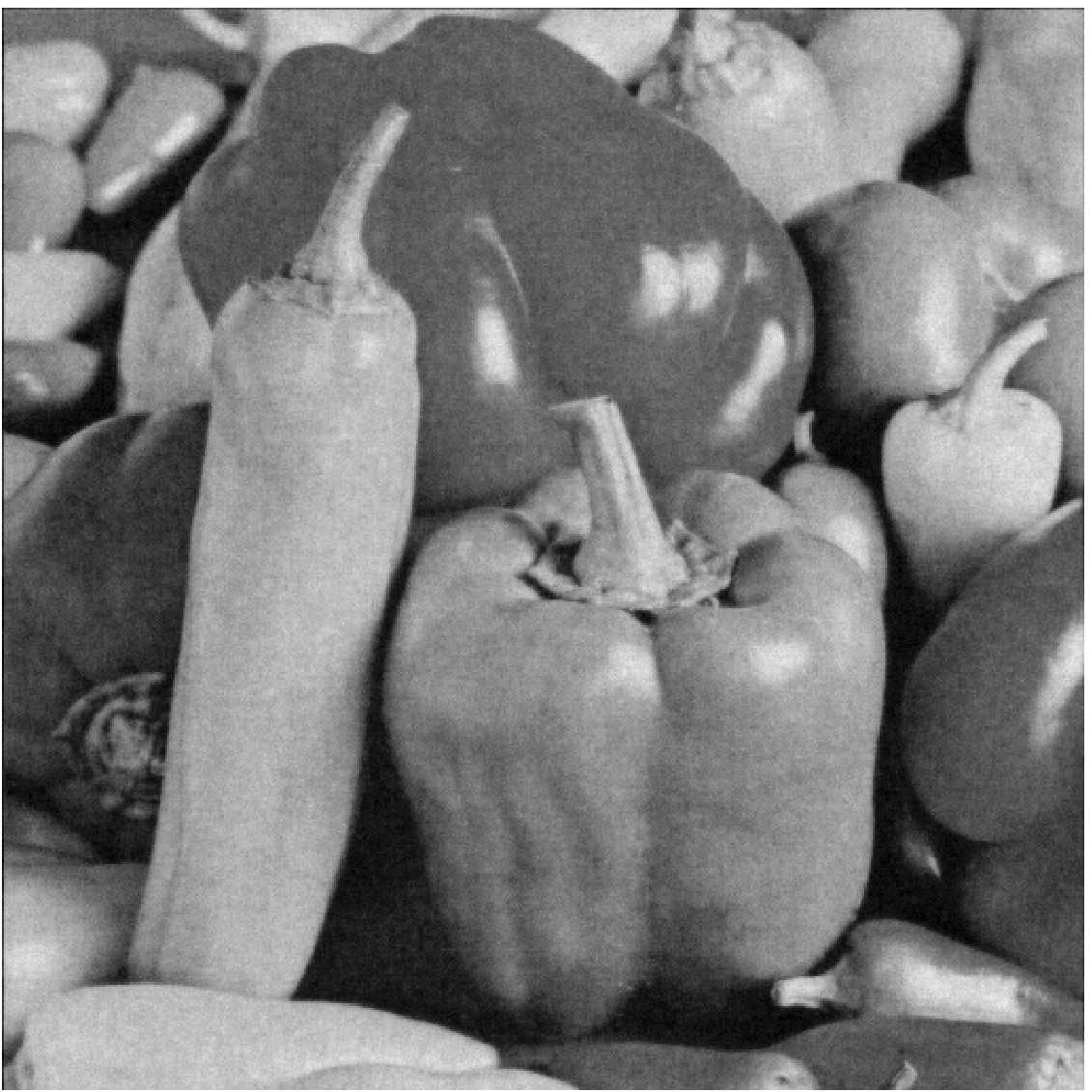}
\end{minipage}}%
\subfigure[]{
\begin{minipage}[t]{0.2\linewidth}
\includegraphics[width=\textwidth]{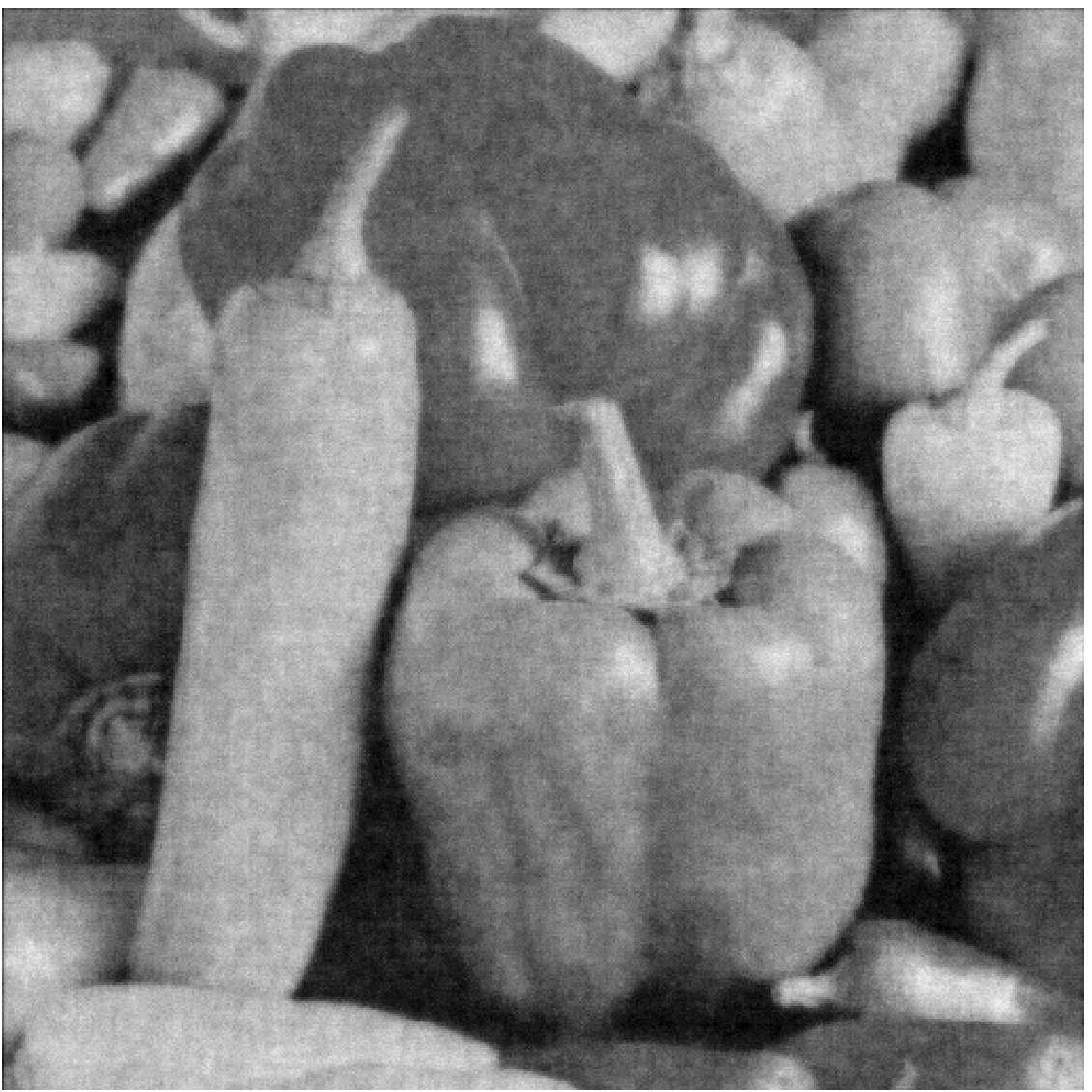}
\end{minipage}}
\caption{ Four decoded images corresponding to the four encoded images in Fig. \ref{fig3} for different $CR$s.  (a) $CR = 0.8$; (b) $CR = 0.6$; (c) $CR = 0.4$; (d) $CR = 0.2$.}
\label{fig5}
\end{figure*}

\begin{figure*}[th]
\centering
\subfigure[]{
\begin{minipage}[t]{0.2\linewidth}
\includegraphics[width=\textwidth]{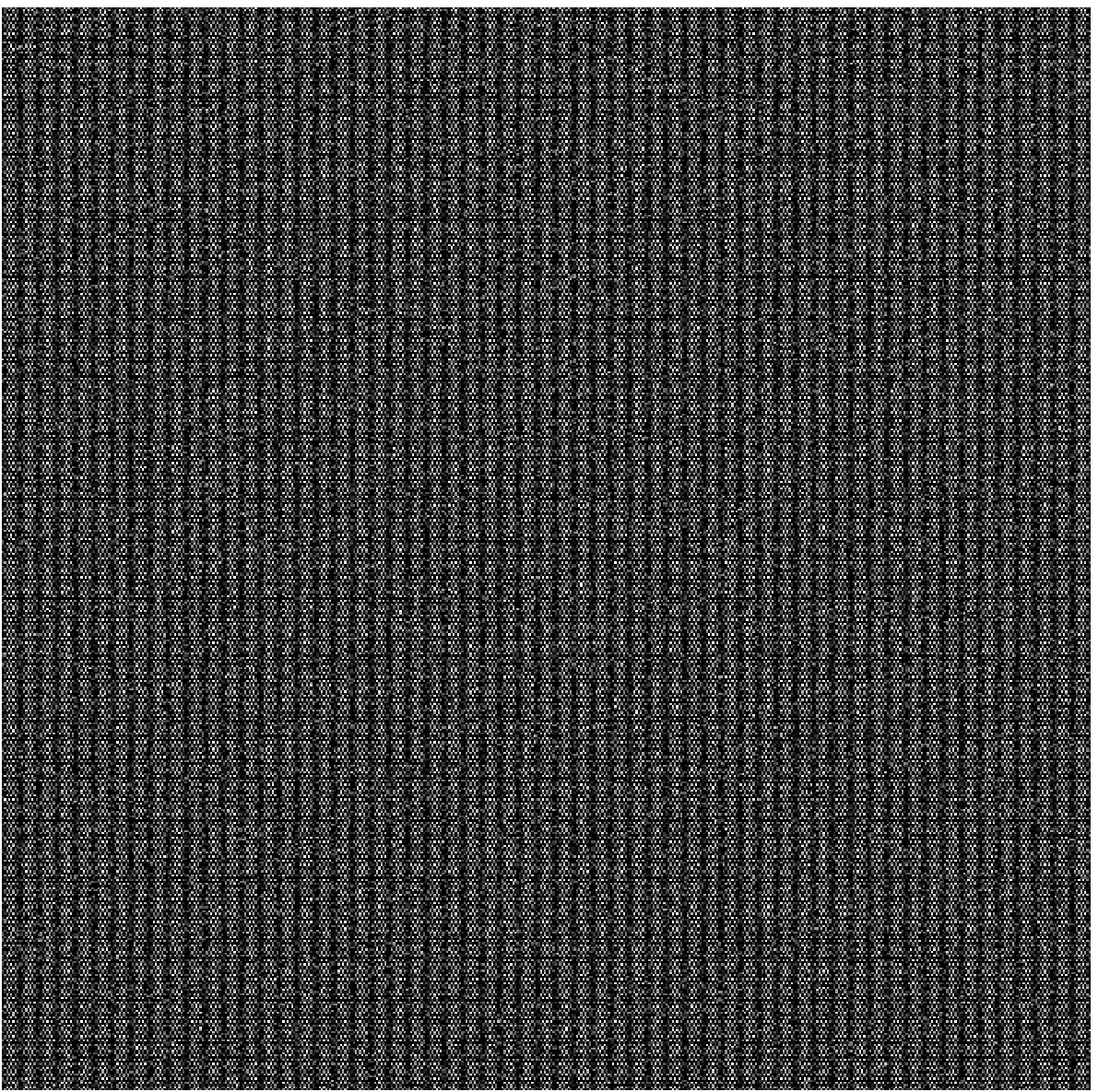}
\end{minipage}}%
\subfigure[]{
\begin{minipage}[t]{0.2\linewidth}
\includegraphics[width=\textwidth]{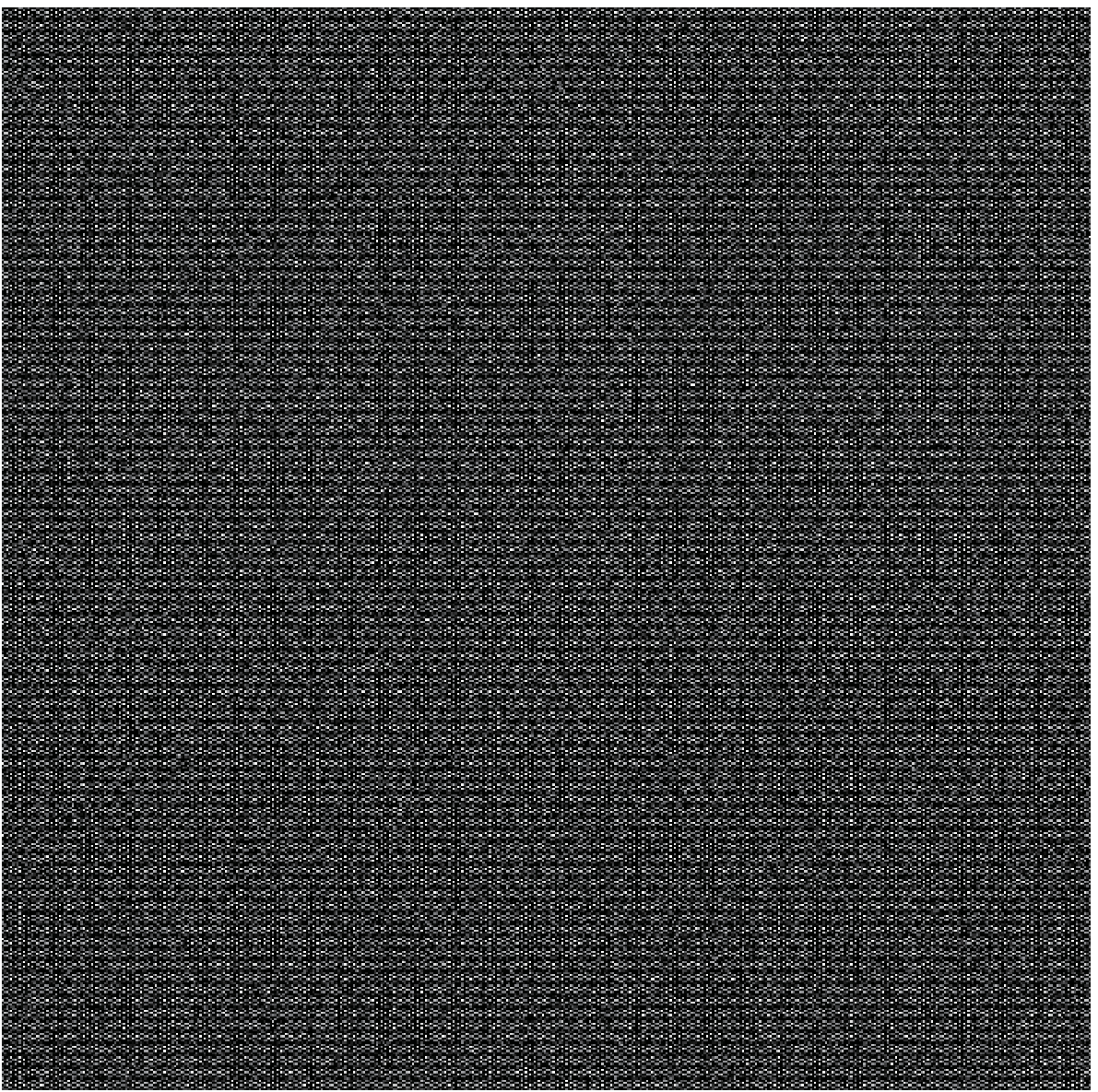}
\end{minipage}}
\subfigure[]{
\begin{minipage}[t]{0.2\linewidth}
\includegraphics[width=\textwidth]{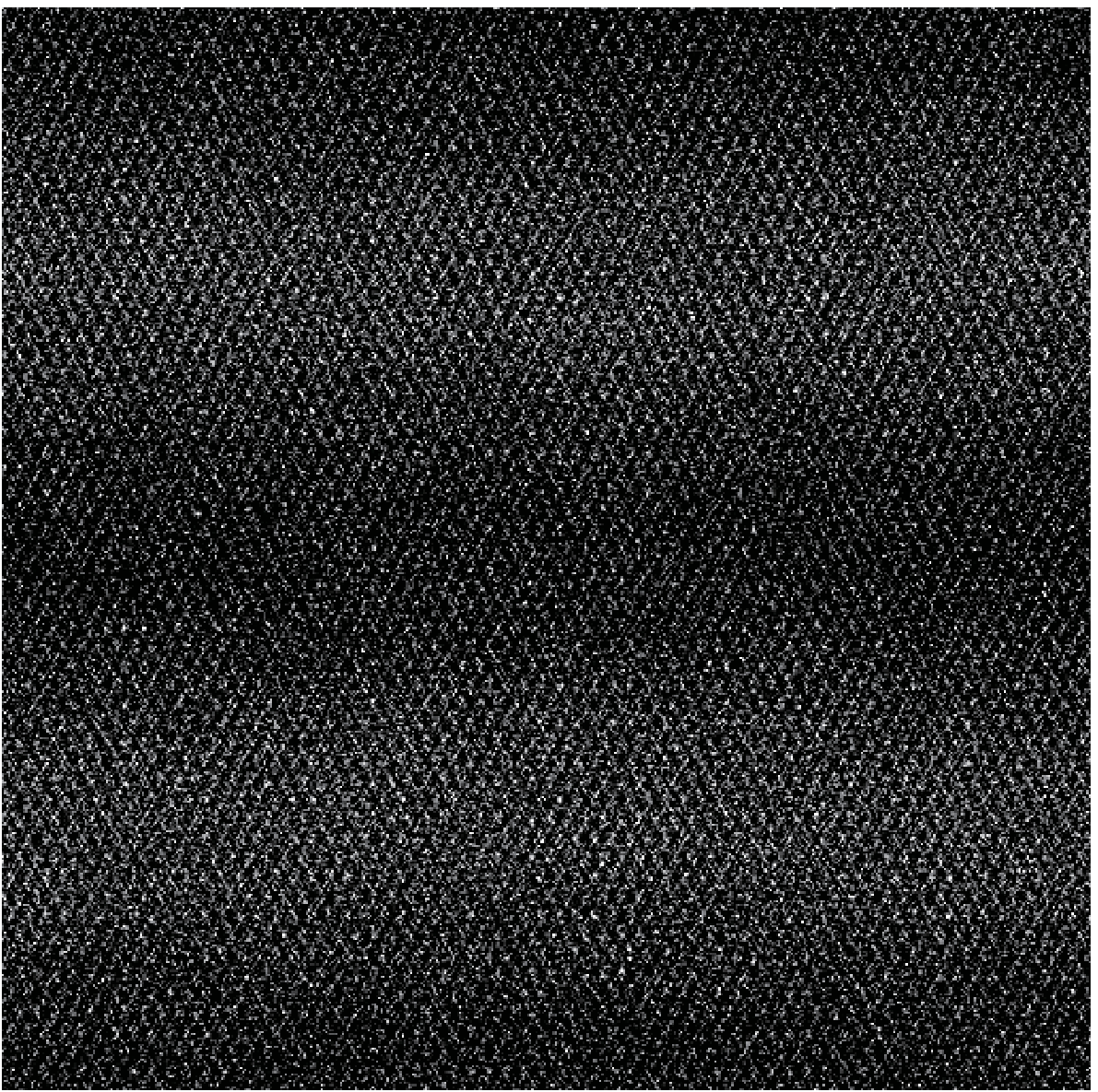}
\end{minipage}}%
\subfigure[]{
\begin{minipage}[t]{0.2\linewidth}
\includegraphics[width=\textwidth]{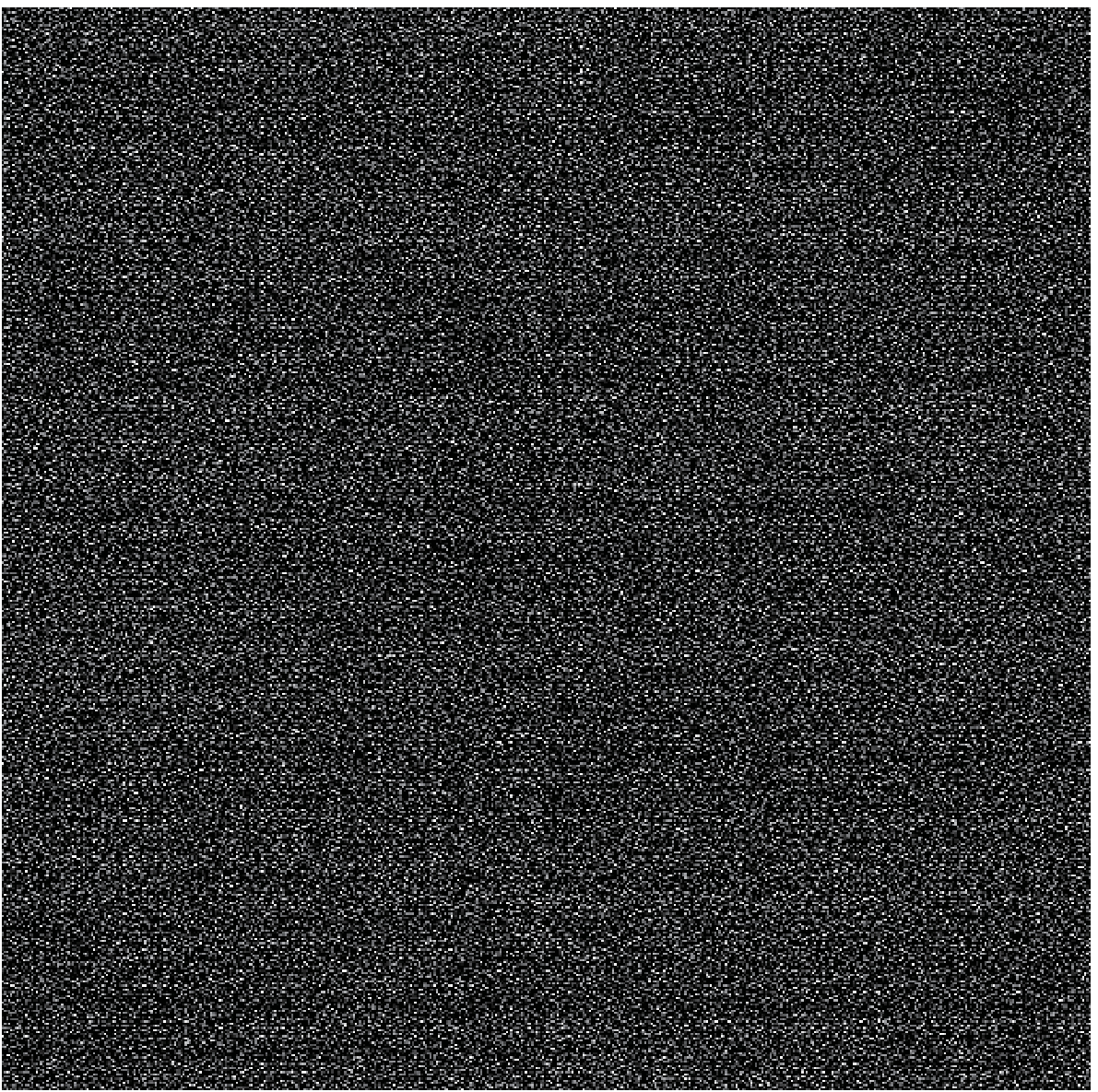}
\end{minipage}}
\caption{ The test of key sensitivity ($CR = 0.2$).  (a) $z\left( 0 \right) = 0.33 + {10^{ - 16}}$; (b) $\mu  = 0.63 + {10^{ - 16}}$; (c) $z'\left( 0 \right) = 0.73 + {10^{ - 16}}$; (d) $\mu ' = 0.28 + {10^{ - 16}}$.}
\label{fig6}
\end{figure*}

\section{Embedding Bi-layer Encryption in Parallel Compressed Sensing using Chaos}

A block diagram of our approach is depicted in Fig. \ref{fig2}. Compressive sensing and encryption are performed simultaneously under the control of chaos. The encoding process is mainly comprised of two steps, permutation and measurement. A 2D signal ${\bf{X}} \in \mathbb{R}^{M \times N}$ is firstly reshaped into 1D signal $\left\{ {x\left( i \right)} \right\}_{i = 1}^{MN}$, which is then permuted in accordance with $\left\{ {Index\left( i \right)} \right\}_{i = 1}^{MN}$ described below. The permuted signal $\left\{ {{x^ * }\left( i \right)} \right\}_{i = 1}^{MN}$ is converted back to the 2D format ${{\bf{X}}^ * } \in \mathbb{R}^{M \times N}$. After the permutation, the signal ${{\bf{X}}^ * }$ is sampled column by column using the same measurement matrix ${\bf{\Phi}}$, i.e., ${{\bf{Y}}^ * }\left[ j \right] = {\bf{\Phi}}{{\bf{X}}^ * }\left[ j \right]$, where ${{\bf{Y}}^ * } \in \mathbb{R}^{K \times N}$ and ${{\bf{X}}^ * }\left[ j \right]$ represents the $j$th column of ${{\bf{X}}^ * }$. In the decoding phase, ${{\bf{\hat X}}^ * }$ can be recovered from the received ${{\bf{\hat Y}}^ * }$ and is then processed by the reverse permutation to derive the signal ${\bf{\hat X}}$ of interest, as shown in Fig. 2. The whole process is controlled by chaos, specifically, the skew tent map with four keys, $\mu $, $z\left( 0 \right)$, $\mu '$ and $z'\left( 0 \right)$.

\subsection{Generating the Permutation Order}

There are a number of classic methods in realizing the permutation operation ${\bf{P}(\bullet)}$ from one or more keys using chaos, some of which are stated as follows:

\emph{Straightforward transform}. Use 2D chaotic transforms such as Arnold map to directly project the indices of the 2D signal, e.g., \cite{zhang2004t}.

\emph{Matrix rotation}. Employ the chaotic sequence to construct the rotation matrix acting on 1D signal, e.g., \cite{zhang2014image}.

\emph{Indices sorting}. Sort the chaotic sequence to generate the index matrix, applying to the  indices of the 1D signal, e.g., \cite{wang2012novel}.

Here, we apply the method ``indices sorting'' to perform the permutation. According to \cite{wang2012novel}, a permutation sequence is produced by using the skew tent map by the following steps:

a. Set the keys $\mu $ and $z\left( 0 \right)$ to iterate the skew tent map $MN + m$ times, then discard the first $m$ values to get rid of the transient effect.

b. Sort the remaining $MN$ values $\left\{ {z\left( i \right)} \right\}_{i = m + 1}^{m + MN}$ to obtain $\left\{ {\bar z\left( i \right)} \right\}_{i = m + 1}^{m + MN}$.

c. Search the values of $\left\{ {z\left( i \right)} \right\}_{i = m + 1}^{m + MN}$ in $\left\{ {\bar z\left( i \right)} \right\}_{i = m + 1}^{m + MN}$, and store the corresponding indices $\left\{ {Index\left( i \right)} \right\}_{i = 1}^{MN}$.

Apparently, $\left\{ {Index\left( i \right)} \right\}_{i = 1}^{MN}$ indicates an order of the integers from 1 to $MN$. The above steps have been widely used to generate the permutation order in image encryption schemes. However, the complexity ${\rm \mathcal{O}}\left( {{n^2}\log n} \right)$ is high. A novel algorithm, whose complexity is reduced to ${\rm \mathcal{O}}\left( {n\log n} \right)$, was designed in \cite{zhang2013chaotic}. The procedures are:

a. Initialize a flag sequence $\left\{ {flag\left( k \right)} \right\}_{k = 1}^{MN}$ and a permutation sequence $\left\{ {Index\left( k \right)} \right\}_{k = 1}^{MN}$
to 0 and set $i = 1$, $j = 1$.

b. Calculate $z\left( {k + 1} \right) = T\left[ {z\left( k \right);\mu } \right]$ and $\chi  = \left\lceil {MN \times z\left( {k + 1} \right)} \right\rceil $.

c. If $flag\left( \chi  \right) = 0$, then set $Index\left( i \right) = \chi$, $flag\left( \chi  \right) = 1$ and $i = i + 1$; otherwise, let $j = j + 1$ and go to Step b.

d. If $i < MN$, set $j = j + 1$ and go to Step b.

\subsection{Generating the Measurement Matrix}

Following the idea of \cite{frunzete2011compressive}, the chaotic measurement matrix is constructed by the following steps:

a. Define the chaotic sequence $\Delta \left( {d,k,\mu ',z'\left( 0 \right)} \right): = \left\{ {z'\left( {n + i \times d} \right)} \right\}_{i = 0}^k$, which is extracted from the chaotic sequence generated by the skew tent map with sampling distance $d$ and keys $\mu '$ and $z'\left( 0 \right)$.

b. Introduce a new transform $\left\{ {\vartheta \left( k \right)} \right\}_{k = 0}^{KM - 1} = {\left. {\left\{ { - 2 \times \Delta \left( {d,k,\mu ',z'\left( 0 \right)} \right) + 1} \right\}} \right|_{k = KM - 1}}$.

c. Create the measurement matrix column by column using the sequence $\left\{ {\vartheta \left( k \right)} \right\}_{k = 0}^{KM - 1}$, as given by

\begin{equation}
{\bf{\Phi}} = \sqrt {\frac{2}{K}} \left( {\begin{array}{*{20}{c}}
   {\vartheta \left( 0 \right)} &  \cdots  & {\vartheta \left( {KM - K} \right)}  \\
    \vdots  &  \ddots  &  \vdots   \\
   {\vartheta \left( {K - 1} \right)} &  \ldots  & {\vartheta \left( {KM - 1} \right)}  \\
\end{array}} \right)
\end{equation}
where the scalar $\sqrt {{2 \mathord{\left/ {\vphantom {2 K}} \right. \kern-\nulldelimiterspace} K}}$ is used for normalization.

\section{Simulation}

An image can be considered as a 2D signal, which is sparsified by 2D discrete cosine transform (DCT2) to obtain a 2D sparse signal ${\bf{X}}$. The best $s$-term approximation is acquired by keeping the $s$ largest DCT2 coefficients and setting the remaining to zeros. Four images of size 512$\times$512, Peppers, Lena, Boat and Baboon, are used in the simulations. The basis pursuit algorithm in the CVX optimization toolbox \cite{grant2008cvx} is employed to realize the PCS reconstruction. Apart from the basis pursuit, other reconstruction algorithms can also be used. The reconstruction performance is evaluated by peak signal-to-noise ratio (PSNR). Four session keys $\mu $, $z\left( 0 \right)$, $\mu '$ and $z'\left( 0 \right)$ are chosen from $\left( {0,1} \right)$, satisfying $\mu  \ne z\left( 0 \right)$ and $\mu ' \ne z'\left( 0 \right)$. With respect to the construction of the measurement matrix, the sampling distance is chosen as $d = 15$ according to \cite{frunzete2011compressive}.

\subsection{Compressibility}

\begin{figure}[th]\centering
 \includegraphics[width=9 cm]{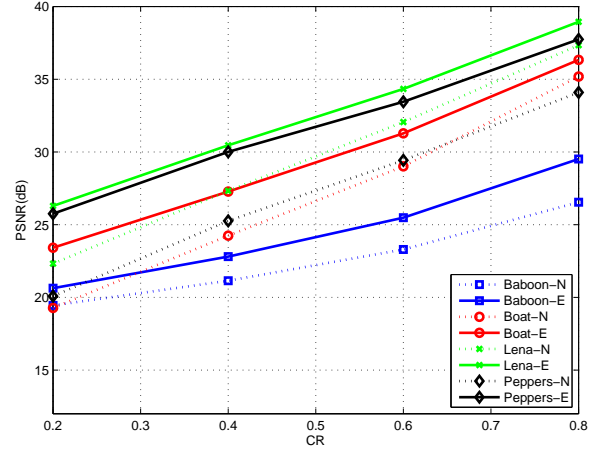}\\
  \caption{ PSNR versus $CR$ for different images with/without encryption.}
  \label{fig4}
\end{figure}

The encoded (or encrypted) image can have various sizes depending on the compression ratio ($CR$), i.e., the ratio of the number of measurements to the total number of entries in the DCT2 coefficient matrix. Figure \ref{fig3} shows four encoded images for the original Peppers image corresponding to $CR = 0.8,0.6,0.4,0.2$ with given key values $\mu  = 0.63$, $z\left( 0 \right) = 0.33$, $\mu ' = 0.28$ and $z'\left( 0 \right) = 0.73$. In order to investigate the effect of encryption on $CR$, we plot PSNR versus $CR$  for different images with/without encryption in Fig. \ref{fig4}, where ``E'' represents encryption while ``N'' means no encryption. No encryption refers to the case that a 2D sparse signal is sampled column by column with the same measurement matrix drawn from Gaussian ensembles, which are replaced by the chaotic sequences in our experiment. As observed from Fig. \ref{fig4}, encryption helps to improve the PSNR of all images by around 2$\sim$6 dB at the same $CR$. In other words, at the same PSNR, encryption makes $CR$ smaller. In comparison with the general joint compression and encryption schemes in which encryption always compromises the compression performance, the proposed approach embedding encryption in PCS reduces $CR$, while maintains the same reconstruction quality. This is mainly due to the fact that random permutation can relax the RIP for 2D sparse signals with high probability in PCS, as justified by Theorem 1.

\subsection{Robustness}

Introducing encryption into PCS makes it still possess high reconstruction robustness, even for a small amount of encoded data. This can be visually verified by the four decoded images, as shown in Fig. \ref{fig5}. The decoded (or decrypted) images contain most of the visual information of the original images, even at $CR = 0.2$. A significant requirement in the transmission process is the robustness of a coding system (or cryptosystem) against imperfection such as additive Gaussian while noise (AGWN) and cropping attack (CA). These two capabilities are quantified in Table \ref{tab2} for the Peppers image. In particular, the encoded images at different $CR$s are affected by these imperfections and the PSNRs of the corresponding decoded images are calculated. Additive Gaussian while noise yields zero-mean normal distribution with variance 1 while the cropping attack cuts one-eighth of the encoded image at the upper left corner. Observing ${\Gamma _3}$ and ${\Gamma _4}$, or ${\Gamma _5}$ and ${\Gamma _6}$ from Table \ref{tab2}, we can see that at a channel with both AGWN and CA, encryption improves the PSNR at the same $CR$. By individually comparing the variation trends of ${\Gamma _2} - {\Gamma _1}$, ${\Gamma _4} - {\Gamma _3}$ and ${\Gamma _6} - {\Gamma _5}$, the tendency is that the smaller the $CR$, the greater the improvement. In addition, vertically contrasting these three rows of data reveals that PSNE improvements are similar with and without AGWN. With respect to CA, the improvement is reduced. Thus, we come to the conclusion that the proposed approach possesses a stronger robustness against AGWN than CA. It is worth mentioning that similar results are obtained using other images.

\begin{table}[tb]
\caption{PSNR of different settings at various $CR$s.}
\centering
\begin{tabular}{l c c c c}
\hline
$CR$ & 0.2 & 0.4 & 0.6 & 0.8\\
\hline
PCS-N(=${\Gamma _1}$) & 20.0800 & 25.2575 & 29.4243 & 34.0949  \\

PCS-E(=${\Gamma _2}$) & 25.7507 & 30.0071 & 33.4539 & 37.7431  \\

${\Gamma _2} - {\Gamma _1}$ & 5.6707 & 4.7496 & 4.0296 & 3.6482  \\

PCS-AGWN-N(=${\Gamma _3}$) & 20.0899 & 25.2382 & 29.3072 & 33.0931  \\

PCS-AGWN-E(=${\Gamma _4}$) & 25.7405 & 29.9143 & 33.0547 & 35.8942  \\

${\Gamma _4} - {\Gamma _3}$ & 5.6506 & 4.6761 & 3.7475 & 2.8011  \\

PCS-CA-N(=${\Gamma _5}$) & 13.7827 & 13.2879 & 11.8456 & 9.0193  \\

PCS-CA-E(=${\Gamma _6}$) & 16.1730 & 15.9257 & 13.7208 & 9.5178  \\

${\Gamma _6} - {\Gamma _5}$ & 2.3903 & 2.6378 & 1.8752 & 0.4985 \\

\hline
\end{tabular}
\label{tab2}
\end{table}

\section{Security Analysis}

In this section, we investigate the security of the proposed scheme embedding bi-layer encryption in PCS using chaos. Assume that Alice sends an encrypted message ${{\bf{Y}}^ * } = {\bf{\Phi P}}\left( {\bf{X}} \right) = {\bf{\Phi }}{{\bf{X}}^ * }$ to Bob, who decrypts the message by solving the following convex optimization problem

\begin{equation}
\min {\left\| {{{\bf{X}}^ * }\left[ j \right]} \right\|_1}\; s.t.\; {{\bf{Y}}^ * }\left[ j \right]{\bf{ = \Phi }}{{\bf{X}}^ * }\left[ j \right], \;j \in \left[ {1,N} \right]
\end{equation}
and so ${\bf{X}} = {{\bf{P}}^{ - 1}}\left( {{{\bf{X}}^ * }} \right)$. An eavesdropper, Eve, attempts to recover the plaintext ${\bf{X}}$ or the encryption keys ${\bf{\Phi }}$ and ${\bf{P}(\bullet)}$ determined by the initial values of the chaotic system after intercepting the ciphertext ${{\bf{Y}}^ * }$.

\subsection{Asymptotic Spherical Secrecy}

Considering Shannon's definition of perfect secrecy that the probability of a message conditioned on the cryptogram is equal to the \emph{a priori} probability of the message, the proposed scheme does not achieve perfect secrecy, as stated in Lemma 3.

\noindent
\textbf{Lemma 3:} Let $X$ be a random varible, whose probability is ${P_X}\left( {\bf{X}} \right) > 0$, $\forall {\kern 1pt} {\kern 1pt} {\bf{X}} \in {\mathbb{R}^{M \times N}}$, and ${\bf{\Phi }}$ be a $K \times M$ measurement matrix. With respect to the encryption model $Y = {\bf{\Phi P}}(X)$, we have $I\left( {X;Y} \right) > 0$, and so perfect secrecy is not achieved.

The proof is given in Appendix B.

By the RIP, ${\bf{Y}}$ provides information about the norm of ${\bf{X}}$. The fact that the ${l_2}$-norm of a vector can be considered as its energy has been utilized by Cambareri \emph{et al.} \cite{cambareri2013low} in introducing the notion of asymptotic spherical secrecy for CS encoding in which the measurement matrix serves as a key.

\noindent
\textbf{Definition 3:} \cite{cambareri2013low} (asymptotic spherical secrecy). Let ${{\bf{x}}^{\left( n \right)}} = \left( {{x_0},{x_1}, \cdots ,{x_n}} \right) \in {\mathbb{R}^n}$ be a plaintext sequence of increasing length $n$ and ${{\bf{y}}^{\left( n \right)}}$ be the
corresponding ciphertext sequence. Assume that the power of the plaintext is positive and finite, i.e.,

\begin{equation}
{W_{\bf{x}}} = \mathop {\lim }\limits_{n \to \infty } {1 \over n}\sum\nolimits_{k = 1}^n {x_k^2} ,{\kern 1pt} {\kern 1pt} {\kern 1pt} {\kern 1pt} {\kern 1pt} 0 < {W_{\bf{x}}} <  + \infty.
\end{equation}

A cryptosytem is said to have asymptotic spherical secrecy if

\begin{equation}
{f_{{Y^{\left( n \right)}}|{X^{\left( n \right)}}}}\left( {{\bf{y}},{\bf{x}}} \right)\mathop  \to \limits_\mathcal{D} {f_{{Y^{\left( n \right)}}|{W_{\bf{x}}}}}\left( {\bf{y}} \right),
\end{equation}
where $\mathop  \to \limits_\mathcal{D} $ denotes the convergence in distribution as $n \to \infty $.

This definition implies that it is impossible for Eve to infer the plaintext ${\bf{x}}$ but its power from the statistical properties of the random measurements ${\bf{y}}$. Although not achieving perfect secrecy, the proposed scheme satisfies asymptotic spherical secrecy.

\noindent
\textbf{Theorem 3:} (asymptotic spherical secrecy of the proposed scheme)

Let

1) ${{\bf{X}}^{\left( n \right)}} = \left( {{X_{ij}}} \right) \in {\mathbb{R}^{M \times N}}$ be a bounded-value plaintext with power $0 < {W_{\bf{X}}} <  + \infty $, defined as ${W_{\bf{X}}} = \mathop {\lim }\limits_{n \to \infty } {1 \over n}\sum\nolimits_{i = 1}^M {\sum\nolimits_{j = 1}^N {X_{ij}^2} } $, where $n = MN$;

2) ${{\bf{X}}^ * }^{\left( n \right)} = {\bf{P}}\left( {{{\bf{X}}^{\left( n \right)}}} \right) = \left( {X_{ij}^ * } \right) \in {\mathbb{R}^{M \times N}}$ with power ${W_{{{\bf{X}}^ * }}} = \mathop {\lim }\limits_{n \to \infty } {1 \over n}\sum\nolimits_{i = 1}^M {\sum\nolimits_{j = 1}^N {{{\left( {X_{ij}^ * } \right)}^2}} } $;

3) ${{\bf{Y}}^{\left( n \right)}} = \left( {{Y_{ij}}} \right) \in {\mathbb{R}^{K \times M}}$ be the corresponding ciphertext, where ${Y_{ij}} = \sum\nolimits_{k = 1}^M {{\Phi _{ik}}X_{kj}^ * } $.
As $n \to \infty $, we have

\begin{equation}
{Y_{ij}}\mathop  \to \limits_\mathcal{D} N\left( {0,{{M{W_{\bf{X}}}} \mathord{\left/
 {\vphantom {{M{W_{\bf{X}}}} K}} \right.
 \kern-\nulldelimiterspace} K}} \right).
\end{equation}

The derivation of this result can be found in Appendix C.

\subsection{Computational Secrecy}
Cryptosystems relying on computation-secrecy such as RSA are practical and widely used. In contrast to information theoretic secrecy which is an ideal encryption requirement, computational secrecy allows the ciphertext contains the complete or partial plaintext information, which is common. This ensures that for Eve to recover the plaintext from the ciphertext without the correct keys is equivalent to solving a computational problem that is assumed to be extremely difficult (e.g., NP-hard). In the proposed scheme, ${\bf{X}}$ is a 2D sparse signal with sparsity ${\bf{s}}$. If a wrong key $\mu $, $z\left( 0 \right)$, $\mu '$ or $z'\left( 0 \right)$, which is almost identical to the correct key, is used by Eve in attempting to recover ${\bf{X}}$, the result is unsuccessful due to the high key sensitivity. To test the sensitivity of the four keys, a tiny perturbation of ${10^{ - 16}}$ is added, respectively, and the decoded images are depicted in Fig. \ref{fig6}. Their indistinguishability justifies the high key sensitivity of the proposed approach. In fact, this is guaranteed by the inherent property of chaos, i.e., high sensitivity to initial conditions. The key space is at least ${2^{64}}$. Moreover, the unsuccessful recovery of the signal using a wrong key can also be justified by the following theorem.

\noindent
\textbf{Theorem 4:} \cite{rachlin2008secrecy} Let ${\bf{\Phi }}$ and ${\bf{\tilde \Phi }}$ be $K \times M$ matrices with entries generated by different keys. Let ${\bf{x}}$ be $s$-sparse and ${\bf{y}} = {\bf{\Phi x}}$. When $\tilde s \ge s + 1$, the ${l_0}$ or ${l_1}$ optimization used will yield an $\tilde s$-sparse solution with probability one.

On the contrary, once an $s$-sparse solution is obtained using some keys, Eve easily realizes that it must be the correct key. Computational secrecy can be achieved if Eve is computationally bounded; otherwise, the cryptanalysis will succeed. However, in practical applications, the keys should be at least ${2^{64}}$ bits and are updated periodically to resist brute-force attack.


\section{Conclusion}
This paper is firstly dedicated to  the design of TLPM for SCS. Some connections between CS and symmetric-key cipher are analyzed. Some cryptographic features are embedded in CS, as summarized in Table \ref{tab1}. We hope the connections between CS and symmetric-key cipher will lead to a new point of view and stimulate further research in both areas. In the second part of this paper, an encryption scheme for PCS has been proposed. Simulations using images as 2D signals show that at the same compression ratio, encryption improves the PSNR by 2$\sim$6 dB for all images. For a channel suffered from both additive Gaussian white noise and cropping attack, encryption still improves the PSNR when the compression ratio is fixed. It is found that the proposed approach possesses a higher robustness against additive Gaussian white noise than cropping attack. The encryption also possesses high key sensitivity and security. In our further work, we will further explore and design some other SCS schemes by virtue of TLPM.

\appendices
\section{Proof of Theorem 1}
\noindent
\textbf{Proof:} If ${\left\| {{{\bf{s}}^ * }} \right\|_\infty } \le {\left\| {\bf{s}} \right\|_\infty }$, i.e., $\Pr \left\{ {{\bf{P}(\bullet)}{\kern 1pt} {\kern 1pt} is{\kern 1pt} {\kern 1pt} acceptable} \right\} = 0$, meaning that each column of ${\bf{X}}$
tends to have similar sparsity levels, ${\bf{P}(\bullet)}$ does not work. However, such an ${\bf{X}}$ has relaxed the RIP condition for PCS without permutation. Thus, we consider the ${\bf{X}}$ where the distribution of the sparsity level in each column is not sufficiently uniform, which, more importantly, accords with the feature of a nature signal. Each element in ${\bf{X}}$ will be randomly located at any index of ${{\bf{X}}^ * }$, that is, this transition of all the indices from ${\bf{X}}$ to ${{\bf{X}}^ * }$ yields the uniform distribution. Each non-zero element of ${\bf{X}}$ appears in each column of ${{\bf{X}}^ * }$ with equal probability $\frac{1}{N}$. This has a strong resemblance to the classical probability problem of $s$ balls and $N$ boxes. The probability that the random permutation ${\bf{P}(\bullet)}$ is an acceptable permutation is calculated as

\begin{equation}
\begin{array}{l}
 {\mathop{\rm P}\nolimits} \left\{ {{\bf{P}}\left( \bullet \right){\kern 1pt} {\kern 1pt} {\kern 1pt} {\kern 1pt} is{\kern 1pt} {\kern 1pt} acceptable} \right\} = {\mathop{\rm P}\nolimits} \left\{ {{{\left\| {{{\bf{s}}^ * }} \right\|}_\infty } < {{\left\| {\bf{s}} \right\|}_\infty }} \right\} \\
  = 1 - {\mathop{\rm P}\nolimits} \left\{ {{{\left\| {{{\bf{s}}^ * }} \right\|}_\infty } \ge {{\left\| {\bf{s}} \right\|}_\infty }} \right\} = 1 - \sum\nolimits_{k = {{\left\| {\bf{s}} \right\|}_\infty }}^N {{\mathop{\rm P}\nolimits} \left\{ {{{\left\| {{{\bf{s}}^ * }} \right\|}_\infty } = k} \right\}}  \\
  = 1 - \sum\nolimits_{k = {{\left\| {\bf{s}} \right\|}_\infty }}^N {\frac{{C_N^1}}{{{N^k}}}}  = 1 - \frac{N}{{N - 1}}\bullet\frac{1}{{{N^{{{\left\| {\bf{s}} \right\|}_\infty }}}}}\bullet\frac{{{N^{N + 1 - {{\left\| {\bf{s}} \right\|}_\infty }}} - 1}}{{{N^{N + 1 - {{\left\| {\bf{s}} \right\|}_\infty }}}}} \\
  = 1 - \frac{{{N^{N + 1 - {{\left\| {\bf{s}} \right\|}_\infty }}} - 1}}{{\left( {N - 1} \right){N^N}}} \ge 1 - \frac{{{N^{N + 1 - \left\lceil {{s \mathord{\left/
 {\vphantom {s N}} \right.
 \kern-\nulldelimiterspace} N}} \right\rceil }} - 1}}{{\left( {N - 1} \right){N^N}}} \\
  \buildrel\textstyle.\over= 1 - \frac{1}{{{N^{\left\lceil {{s \mathord{\left/
 {\vphantom {s N}} \right.
 \kern-\nulldelimiterspace} N}} \right\rceil }}}}. \\
 \end{array}
\end{equation}
This completes the proof.

\section{Proof of Lemma 3}
\noindent
\textbf{Proof:} We prove this lemma by contradiction. Apparently, $I\left( {X;Y} \right) > 0$ if and only if $X$
and $Y$ are statistically independent. In the context of $X = {\bf{0}}$, $Y = {\bf{\Phi P}}(X) = {\bf{\Phi P}}({\bf{0}}) = {\bf{\Phi }} \cdot {\bf{0}} = {\bf{0}}$ and thus ${P_{Y|X}}\left( {Y = {\bf{0}}|X = {\bf{0}}} \right) = 1$. On the other hand, only ${\bf{X}}$
in the null space of ${\bf{\bar \Phi }}$ which is a new transform ${\bf{\bar \Phi }} = {\bf{\Phi P(\bullet)}}$ are mapped to $Y = {\bf{0}}$; whereas, we have ${P_Y}\left( {Y = {\bf{0}}} \right) < 1$ due to the assumption that ${P_X}\left( {\bf{X}} \right) > 0$, $\forall {\kern 1pt} {\kern 1pt} {\bf{X}} \in {\mathbb{R}^{M \times N}}$. As a result, we conclude that ${P_{Y|X}}\left( {Y = {\bf{0}}|X = {\bf{0}}} \right) \ne {P_Y}\left( {Y = {\bf{0}}} \right)$, meaning that $X$ and $Y$ are statistically dependent.

\section{Proof of Theorem 3}
\noindent
\textbf{Proof:} Permutation does not affect the power and thus ${W_{{{\bf{X}}^ * }}} = {W_{\bf{X}}}$. After the random permutation, the energy is approximately uniformly dispersed to each column of ${{\bf{X}}^ * }^{\left( n \right)}$. In other words, the power of each column converges to that for the whole plaintext in distribution, i.e.,

\begin{equation}
\begin{array}{l}
 \frac{1}{M}\sum\nolimits_{k = 1}^M {{{\left( {X_{ij}^ * } \right)}^2}} \mathop  \to \limits_D \mathop {\lim }\limits_{n \to \infty } \frac{1}{n}\sum\nolimits_{i = 1}^M {\sum\nolimits_{j = 1}^N {{{\left( {X_{ij}^ * } \right)}^2}} }  \\
  = \mathop {\lim }\limits_{n \to \infty } \frac{1}{n}\sum\nolimits_{i = 1}^M {\sum\nolimits_{j = 1}^N {X_{ij}^2} }  = {W_{\bf{X}}}. \\
 \end{array}
\end{equation}
The measurement matrix ${\bf{\Phi }}$ is sub-Gaussican with i.i.d. elements \cite{frunzete2011compressive}. We calculate

\begin{equation}
\begin{array}{l}
 {\bf{E}}\left[ {Y_{ij}^2} \right] = {\bf{E}}\left[ {{{\left( {\sum\nolimits_{k = 1}^M {{\Phi _{ik}}X_{kj}^ * } } \right)}^2}} \right]{\kern 1pt} {\kern 1pt} {\kern 1pt}  \\
  = \frac{1}{K}\sum\nolimits_{k = 1}^M {{{\left( {X_{kj}^ * } \right)}^2}} \mathop  \to \limits_\mathcal{D} \frac{M}{K}{W_{\bf{X}}}. \\
 \end{array}
\end{equation}
thereby yielding the result stated in Theorem 3.

\bibliographystyle{IEEEtr}
\bibliography{ref}

\end{document}